\documentclass[conference]{IEEEtran}
\IEEEoverridecommandlockouts
\usepackage{cite}
\usepackage[numbers]{natbib}
\usepackage{amsmath,graphicx,booktabs,arydshln}
\usepackage{amsmath,amssymb,amsfonts}
\usepackage{algorithmic}
\usepackage{graphicx}
\usepackage{textcomp}
\usepackage{xcolor}
\def\BibTeX{{\rm B\kern-.05em{\sc i\kern-.025em b}\kern-.08em
    T\kern-.1667em\lower.7ex\hbox{E}\kern-.125emX}}
\begin{document}

\title{Regression-based Music Emotion Prediction using Triplet Neural Networks\\
}

\author{\IEEEauthorblockN{Kin Wai Cheuk, Yin-Jyun Luo}
\IEEEauthorblockA{\textit{Information Systems,}\\
	\textit{Technology, and Design}\\
	\textit{Singapore University}\\
	\textit{of Technology and Design}\\\\
	\textit{Institute of High Performance}\\
	\textit{Computing, A*STAR}\\
Singapore\\
kinwai\_cheuk@mymail.sutd.edu.sg\\
yinjyun\_luo@mymail.sutd.edu.sg}
\and
\IEEEauthorblockN{Balamurali B, T}
\IEEEauthorblockA{\textit{Information Systems,}\\
	\textit{Technology, and Design}\\
	\textit{Singapore University}\\
	\textit{of Technology and Design}\\
Singapore \\
balamurali\_bt@sutd.edu.sg}
\and
\IEEEauthorblockN{Gemma Roig}
\IEEEauthorblockA{\textit{Computer Science Department}\\
	\textit{Goethe University}\\
	\textit{Frankfurt am Main}\\
	Germany \\
	roig@cs.uni-frankfurt.de}
\and
\IEEEauthorblockN{Dorien Herremans}
\IEEEauthorblockA{\textit{Information Systems,}\\
	\textit{Technology, and Design}\\
	\textit{Singapore University}\\
	\textit{of Technology and Design}\\\\
	\textit{Institute of High Performance}\\
	\textit{Computing, A*STAR}\\
	Singapore\\
dorien\_herremans@sutd.edu.sg}
}

\maketitle

\begin{abstract}
In this paper, we adapt triplet neural networks (TNNs) to a regression task, music emotion prediction. 
Since TNNs were initially introduced for classification, and not for regression, we propose a mechanism that allows them to provide meaningful low dimensional representations for regression tasks. We then use these new representations as the input for regression algorithms such as support vector machines and gradient boosting machines.  %This technique can be also used as a state-of-the-art dimensionality reduction technique in a regression task with audio.%Previous music emotion recognition research typically uses a large number of acoustic features (over 6000) to train a regression model, which reduces the scalability of the model when the dataset becomes bigger. DH: how does it reduce the scalability? Could you explain? 
%\dorien{Raven: see comment. How is a large number of features bad for scalability?}
To demonstrate the TNNs' effectiveness at creating meaningful representations, we compare them to different dimensionality reduction methods on music emotion prediction, i.e., predicting valence and arousal values from musical audio signals.
%In a comparison between our proposed method and more traditional dimension reduction techniques such as principal component analysis, we can see that TNN outperforms the latter. 
Our results on the DEAM dataset show that by using TNNs we achieve 90\% feature dimensionality reduction with a 9\% improvement in valence prediction and 4\% improvement in arousal prediction with respect to our baseline models (without TNN). Our TNN method outperforms other dimensionality reduction methods such as principal component analysis (PCA) and autoencoders (AE). This shows that, in addition to providing a compact latent space representation of audio features, the proposed approach has a higher performance than the baseline models. %\dorien{compared to...}. 
\end{abstract}


\section{Introduction}
\label{sec:intro}
The link between music and emotions has been investigated extensively by cognitive scientists and musicologists over the years~\cite{pratt1952music, juslin2001music, herremans2016tension, thao2019multimodal}, which has caused the emergence of the field of automatic music emotion recognition (MER). Being able to predict the emotion from a music audio clip has a myriad of applications, such as managing personal music collections~\cite{cunningham2004organizing}, mood-based music recommendation~\cite{han2010music,kuo2005emotion,inskip2012towards,mandel2006support}, musicology~\cite{herremans2017imma, herremans2017multi}, and music therapy~\cite{sourina2012real, huq2009sourcetone}.  %, automatic play-list generation\cite{mandel2006support}, 
%and song searching \cite{inskip2012towards} 
%\dorien{song searching not the same as recommendation? Play list generation is also the same as music recommendation perhaps?}. However, predicting emotion from music remains a challenging task due to the subjective perception of emotion. In practice, the problem is often simplified by using a limited set of discrete emotions. For instance, the annual music information retrieval evaluation exchange (MIREX) \cite{downie20082007} uses five classes of emotion.  %Since human emotion is rich in nature, classifying emotion into
In order to label emotions, researchers often use a two-dimensional arousal-valence (A/V) representation~\cite{russell1980circumplex,thayer1990biopsychology}. 
%In this model, valence represents the mood (either negative or positive), whereas arousal captures the energy of the music. The DEAM dataset uses A/V values as the annotation. 
Given that A/V values are continuous values, we will approach the emotion prediction task as a regression problem. 
%\footnote{http://cvml.unige.ch/databases/DEAM/}

One of the challenges when tackling automatic emotion prediction from audio is to identify the ideal audio features that best capture the emotion evoked by the audio signal~\cite{huq2010automated}. \citet{aljanaki2013mirutrecht} investigated the performance of 44 audio features extracted using MIRToolbox, PsySound, and SonicAnnotator. Out of these 44 features, they found that 26 features were ideal when predicting arousal and 27 features for valence.
%\dorien{note, do we ever refer back to this or do we include these features?}\Raven{No, we didn't use these features in our experiments.} 
In a paper published by \citet{weninger2013tum}, 6,373 features were to train a support vector machine model to predict A/V. Although the number of features used in this study is many times larger than the 44 \cite{aljanaki2013mirutrecht}, the improvement in $R^2$ score for estimating arousal values is marginal (0.65 versus 0.64). The $R^2$ score for valence, however, was found to be relatively better (0.42 versus 0.36)~\cite{weninger2013tum}. The main challenge when using such a large feature space is that it requires a lot of computational resources, or even faces scalability issues, as reported by \citet{markov2013music}. When their model was trained with a lot less data, the performance of the regression was affected severely. An efficient dimensional reduction technique may provide a solution for this issue. In this paper, we tackle the MER problem by adopting a regression version of the triplet neural network structure in order to reduce the dimensionality, while at the same time creating a more meaningful representation, thus improving the result of the prediction compared to other dimensionality reduction techniques. The source code is available online\footnote{https://github.com/KinWaiCheuk/IJCNN2020\_music\_emotion}.

\section{Regression Using a TNN}
\subsection{Triplet neural networks}
\begin{figure*}[h]
	\centering
	\includegraphics[width=0.85\textwidth]{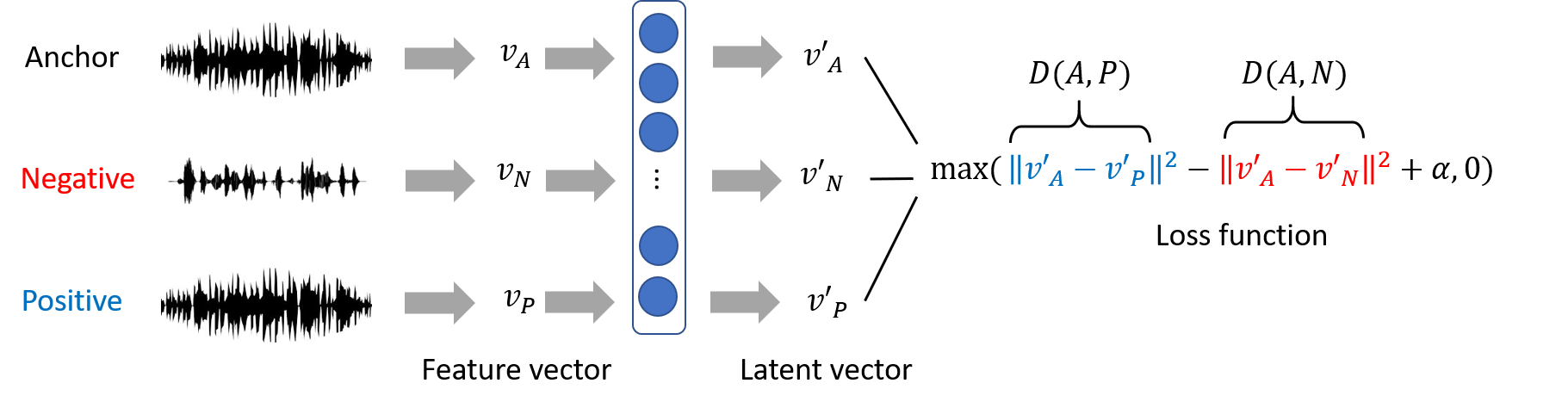}
	\caption{Schematic diagram of the triplet neural network.} 
	%The input of the network is a triplet that consists of anchor, negative, and positive vectors while the output is the latent vector. In the training phase, the output vectors are used to minimize the triplet loss. After training, the network is used to transform vectors to its latent space. \dorien{Great picture, just wondering if we should make it clear that they are fed to the same NN. Is the box clear or should we have three boxes?}}\Raven{If we use three boxes, some people might misunderstand that we uses three different NN for each input. That's why I put only 1 NN here.}
	% \label{fig:my_label}
	\label{fig:training}
\end{figure*}

TNN is a neural network technique first proposed by researchers from Google in 2015~\cite{schroff2015facenet}. Since then it has been used for a variety of tasks such as face recognition or person identification~\cite{chen2016deep,cheng2016person,su2016deep,wang2016joint,ding2015deep, cheuk2019latent}. 
Originally, TNNs were used in classification problems to learn a new feature representation of data, so that this new representation can easily be disentangled with regard to the classes. TNNs consist of 3 inputs, namely anchor ($A$) input, positive ($P$) input, and negative ($N$) input. The positive input belongs to the same class as that of the anchor input while the  negative input is from a different class than the anchor input. This triplet data is fed to a neural network which shares the same weights for each of these 3 inputs. The output of the TNN is then passed to a triplet loss function $L(A,P,N)$ as shown below. %Equation~\eqref{eq:distance}, 
\begin{equation}\label{eq:distance}
L(A,P,N) = \max(D(A,P) - D(A,N) + \alpha, 0) ,
\end{equation}
where $D(A,P)$ represents the squared Euclidean distance between the anchor vector $v'_A$ and the positive example vector $v'_P$;  $D(A,N)$ represents the squared Euclidean distance between $v'_A$ and the negative example vector $v'_N$ as defined in Figure~\ref{fig:training}; and $\alpha$ is the margin parameter that specifies how far positive and negative examples should be apart. The training goal is to reduce the distance  between similar samples $D(A,P)$, and increase the distance for samples in different classes $D(A,N)$. In other words, we want to learn a representation that positions vectors that belong to the same class close together and those from a different class far apart. Note that the max operation in Eq.~\eqref{eq:distance} is equivalent to the ReLu operator.
% \begin{comment}
% \end{comment}
\subsection{Defining positive and negative samples for regression}\label{sec: define}
When dealing with a classification task, the definition of positive (or negative) samples is straightforward, i.e., those belonging (or not belonging) to the same class. 
%For example, in the MNIST dataset, a dataset that contains images of handwritten digits and their corresponding digit label (ten possible classes (0-9)). If we pick an image that belongs to the class `4' as an anchor, then all other samples that belong to class `4' will be the positive samples for this anchor. The negative samples for this anchor, on the other hand, consist of all samples from class `0' to class `9' excluding class `4'. 
In the case of regression, however, a new strategy is needed because it operates in the continuous space. Instead of a finite number of discrete classes, the labels now have an infinite number of possible values even in the range $[-1,1]$. Despite attempts in applying TNNs~\cite{lu2017deep,yang2018triplet} to regression problems, this only works if the dataset is specially designed for this type of task. In other words, regression methods cannot be generalized to other datasets unless they are tailored to a regression task.  For example, in \citet{yang2018triplet}'s research, the positive and negative samples are defined by using two classes (pre-triage: before seeing the doctor; and post-triage: after seeing the doctor) of videos without using the true annotation (blood pressure) of the dataset. Without the explicit information about pre-triage and post-triage, the method would not work. \citet{lu2017deep}'s dataset already includes a label for the positive and negative samples, so they can directly form the triplets without any triplet mining. Although they successfully apply TNN to a regression problem, their method only works for this specific (labeled) dataset, and cannot be applied to our problem. For instance, if we have an anchor with value 0.1, should the sample with value 0.3 be a positive sample, or a negative sample? Setting absolute, discrete bins does not work well for regression. For example, let us define the interval $[0,0.5)$ to be bin A, and $[0.5,1]$ to be bin B.  For the anchor sample $y_{v_a}=0.5$, we need to decide if the two samples $y_{v_1}=0.4$ and $y_{v_2}=1$, are positive or negative samples. Under the discrete absolute binning scheme, $y_{v_1}$ will be a negative sample of the anchor since they belong to different bins, while $y_{v_2}$ will be a positive sample of the anchor since they are in the same bin. This could skew our results, as $y_{v_1}$ is obviously closer to the anchor than $y_{v_2}$. Therefore, we propose  a more effective method to define positive and negative samples in regression below. In this section, we describe our mechanism for defining positive and negative samples without explicit information on positive and negative examples.%\dorien{suggestion, if you really want to cut text, I think we could remove the following binning example} One option would be to bin the valence and arousal values into discrete classes, however, the model would not be able to learn an effective latent space for the vector. To alleviate this issue, we propose the following definition for positive and negative samples in a regression problem.

%For example, we bin the A/V values into two classes $A$ and $B$, where $A \in [0, 1]$ and $B\in[1, 2]$. Suppose we have an anchor vector with valence value $0.9$ (i.e. class $A$), and two vectors with valence value $0$ and $1.1$. Under this absolute binning scheme, the vector with valence value $0$ would be defined as positive, while the vector with valence value $1.1$ would become negative. Yet, the anchor having a valence value of $0.9$ is closer to the valence value $1.1$ rather than $0$. Therefore, a simple binning method fails to define an effective positive and negative sample. To alleviate this issue, we propose the following definition for positive and negative samples in a regression problem.

\begin{figure}[h]
	\centering
	\includegraphics[width = 0.32\textwidth]{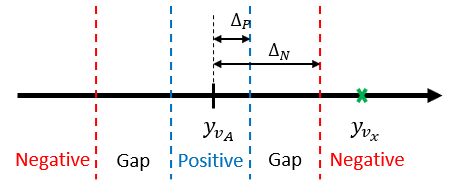}
	\caption{
		%The vector is defined as either positive or negative depending on which region it falls in relative to the anchor's label position. 
		Defining anchor, positive and negative samples. If the sample lies in the gap region, it will discarded and it will not be used for training.}
	% \label{fig:my_label}
	\label{fig:define_PN}
\end{figure}

 In our sampling approach, instead of using a fixed binning, we define fixed threshold gaps that are applied on the anchor samples. Let the valence value for the anchor vector be $y_{v_A}$.  Let $\Delta_P$ be the threshold value, such that for a vector with valence value $y_{v_X} \in [y_{v_A}-\Delta_P, y_{v_A}+\Delta_P]$, the sample is considered to be a positive example. Similarly, we define another threshold value $\Delta_N$, such that the vector $y_{v_X}$ is considered to be a negative sample if $y_{v_X} \in (-\infty, y_{v_A}-\Delta_N] \lor  [y_{v_A}+\Delta_N, \infty)$. The difference between $\Delta_N$ and $\Delta_P$ form a `gap' that guides the network to learn the more distinguishing samples first. The same process is used to define positive and negative arousal values. A schematic representation of the positive and negative sample definition is shown in Figure~\ref{fig:define_PN}. In our experiments, we normalize arousal and valence values so that they fall within $[-1,1]$. The threshold values should be adjusted according to the data distribution. In our experiments, we set $\Delta_P=0.1$ and  $\Delta_N=0.5$, which has been found to perform best through cross-validation.

\section{Dimensionality reduction for emotion prediction}
\label{sec:experiment}

Most research on static emotion annotation (i.e., one rating per song) is based on the 2013 MediaEval dataset\footnote{\label{foot:dataset2013}http://cvml.unige.ch/databases/emoMusic/} (a subset of the DEAM dataset)\cite{fukuyama2016music,weninger2013tum,aljanaki2013mirutrecht,markov2013music,1000SongforEmotioninMusic,openSMILE}. We therefore compare our results with prominent papers using this dataset, and refer to this as the MediaEval experiment. We then also use the complete DEAM dataset\footnote{http://cvml.unige.ch/databases/DEAM/} to test our model, and refer to this as the DEAM experiment. We implement our novel TNN-regression approach for dimensionality reduction and combine it with both a support vector regressor (SVR) and gradient boosting machine~(GBM) to solve the regression problem for the valence and arousal values.

\subsection{MediaEval experiment}
For this experiment we use the 2013 MediaEval dataset\footnotemark[2]. It contains a total of 744 audio files with 6,669 features (listed in Table~\ref{tab:opensmile6669}). Each audio file is labeled with A/V values \cite{schuller2013interspeech}. Different statistical features such as maximum, minimum, mean, range, and etc. are included in the dataset for each of the base features, resulting in 6,669 features in total. 
\begin{table}[htbp]
	\caption{List of openSMILE features provided by the MediaEval 2013 dataset. Different summary statistic features are provided for these base features, resulting in a total of 6,669 features~\cite{soleymani2013mediaeval}}
	\begin{center}
		\begin{tabular}{lr}\toprule
			Feature Name                     & Size \\\midrule
			F0                            & 117     \\ 
			F0env                         & 117     \\
			mfcc                          & 1,521    \\
			pcm\_LOGenergy                & 117     \\
			pcm\_Mag\_fband0-250          & 117     \\
			pcm\_Mag\_fband0-650          & 117     \\
			pcm\_Mag\_fband1000-4000      & 117     \\
			pcm\_Mag\_fband250-650        & 117     \\
			pcm\_Mag\_fband3010-9123      & 117     \\
			pcm\_Mag\_melspec             & 3,042    \\
			pcm\_Mag\_spectralCentroid    & 117     \\
			pcm\_Mag\_spectralFlux        & 117     \\
			pcm\_Mag\_spectralMaxPos      & 117     \\
			pcm\_Mag\_spectralMinPos      & 117     \\
			pcm\_Mag\_spectralRollOff25.0 & 117     \\
			pcm\_Mag\_spectralRollOff50.0 & 117     \\
			pcm\_Mag\_spectralRollOff75.0 & 117     \\
			pcm\_Mag\_spectralRollOff90.0 & 117     \\
			pcm\_zcr                      & 117     \\
			voiceProb                     & 117     \\ \bottomrule

		\end{tabular}
		\label{tab:opensmile6669}
	\end{center}
\end{table}

\begin{table}[htbp]
\caption{List of openSMILE features when using the configuration file \textbf{IS13\_ComParE\_lld-func.conf}. Different summary statistic features are provided for these base features, resulting in a total of 260 features~\cite{AlajankiEmoInMusicAnalysis,1000SongforEmotioninMusic}.}
	\begin{center}
		\begin{tabular}{lr}\toprule
			Feature Name                     & Size \\\midrule
			F0final                          & 4       \\ 
			audSpec\_Rfilt                   & 104     \\
			audspecRasta\_lengthL1norm       & 4       \\
			audspec\_lengthL1norm            & 4       \\
			jitterDDP                        & 4       \\
			jitterLocal                      & 4       \\
			logHNR                           & 4       \\
			pcm\_RMSenergy                   & 4       \\
			pcm\_fftMag\_fband1000-4000      & 4       \\
			pcm\_fftMag\_fband250-650        & 4       \\
			pcm\_fftMag\_mfcc                & 56      \\
			pcm\_fftMag\_psySharpness        & 4       \\
			pcm\_fftMag\_spectralCentroid    & 4       \\
			pcm\_fftMag\_spectralEntropy     & 4       \\
			pcm\_fftMag\_spectralFlux        & 4       \\
			pcm\_fftMag\_spectralHarmonicity & 4       \\
			pcm\_fftMag\_spectralKurtosis    & 4       \\
			pcm\_fftMag\_spectralRollOff25.0 & 4       \\
			pcm\_fftMag\_spectralRollOff50.0 & 4       \\
			pcm\_fftMag\_spectralRollOff75.0 & 4       \\
			pcm\_fftMag\_spectralRollOff90.0 & 4       \\
			pcm\_fftMag\_spectralSkewness    & 4       \\
			pcm\_fftMag\_spectralSlope       & 4       \\
			pcm\_fftMag\_spectralVariance    & 4       \\
			pcm\_zcr                         & 4       \\
			shimmerLocal                     & 4     \\ \bottomrule
		\end{tabular}
		\label{tab:opensmile260}
	\end{center}
\end{table}

The TNN implemented in this experiment consists of a single fully connected layer with 600 neurons and ReLU as the activation function. When using less neurons than 600, the model performance decays. With more neurons, there are too many features such that the classifiers cannot be trained effectively. The network was first trained using 50,000 triplet pairs sampled from the dataset. The mining of the 50,000 triplet pairs consists of the following steps:

\begin{enumerate}
\item Randomly pick a sample as the anchor $v_A$ from the dataset, which has a label value of $y_{v_A}$.

\item Find the positive sample $v_P$ for the anchor using the method mentioned in~\ref{sec: define}.

\item Find the negative sample $v_N$ for the anchor using the method mentioned in~\ref{sec: define}.

\item A triplet is formed with the result from the previous steps ($v_A$, $v_P$, $v_N$).

\item Repeat steps 1 to 4 by choosing other samples as the anchor until the number of mined triplets has reached $50,000$.

\end{enumerate}

With this mining method, we can obtain 50,000 triplet pairs from the original 744 individual audio files.

After an initial 10 epochs, another 50,000 triplet samples were generated and the process continued. In this way, we prevent the model from overfitting to a small set of triplet samples. In total, this was repeated 25 times (i.e., 250 epochs in total) until the model converges, thereby allowing the network to train on as many samples as possible. An Adam optimizer with learning rate $10^{-5}$ was used to minimize the triplet loss during training.

We compared our proposed method with other models applying to the same dataset. \citet{markov2013music} trained a support vector regressor (SVR) using the features provided in the dataset. Their $R^2$ score, however, are only 0.112 and 0.300 for valence and arousal respectively (see Table~\ref{tab:1})~\cite{markov2013music}. They further reported scaling issues with their model due to the large number of audio features. \cite{fukuyama2016music} modified the GPR model to obtain a greater accuracy.
Our approach also addresses the scalability issue reported by \citet{markov2013music}, by using a TNN to reduce the number of features from 6,669 features to 600. 
Both the original features (baseline model) and the transformed low dimensional vectors were used to train an SVR and GBM (referred to as TNN-SVR and TNN-GBM, respectively).  

To evaluate the ability of learning an effective latent space, we compared the TNN results against other dimensionality reduction techniques such as principal component analysis (PCA), Gaussian random projection (RP), and a neural network-based autoencoder (AE). For PCA and RP,  the number of components in the transformed space  was set to 600 so as to match the number of neurons of the TNN; and the random state for the RP was set to 50. For AE, 600 neurons with ReLU activation were used (because our TNN model also uses 600 neurons) for the encoder, and the auto-encoder was trained for until convergence (around 100 epochs). We used standard $R^2$ scores with ten fold cross-validation. For each fold, the weights of the TNN were reset and retrained.

\subsection{DEAM experiment}

The setup for this experiment is similar to that of the MediaEval experiment, except that we use a different dataset with different TNN configuration. The state-of-the-art comparison models remain the same. The DEAM dataset contains 1,724 songs  \cite{AlajankiEmoInMusicAnalysis,1000SongforEmotioninMusic} labeled with A/V values. The dataset provides a total of 260 features extracted from the audio clips using the configuration file \textbf{IS13\_ComParE\_lld-func.conf}. The list of features extracted with the configuration file is listed in Table~\ref{tab:opensmile260}.  For each of the feature, different statistics are calculated such as means, standard deviations, first derivatives, and different bands (for MFCC and specRfilt), resulting in 260 openSMILE features in total. The objective of this experiment is to show the TNN's ability of perform efficient dimensionality reduction on a larger dataset.

We focus on learning a latent space with TNNs and use the learned latent space as the model input for the SVM and GBM regressor. Two different sets of TNN parameters were tested in this experiment: 1) 100 neurons (around one half of the original features); and 2) 50 neurons (around one fourth of the original features) in the fully connected layer. The training procedure is same as in the MediaEval experiment except that 150,000 samples were generated each time. Other parameters are kept the same as in the MediaEval experiment.

\begin{figure}[h]
	\centering
	\includegraphics[width=.4\textwidth]{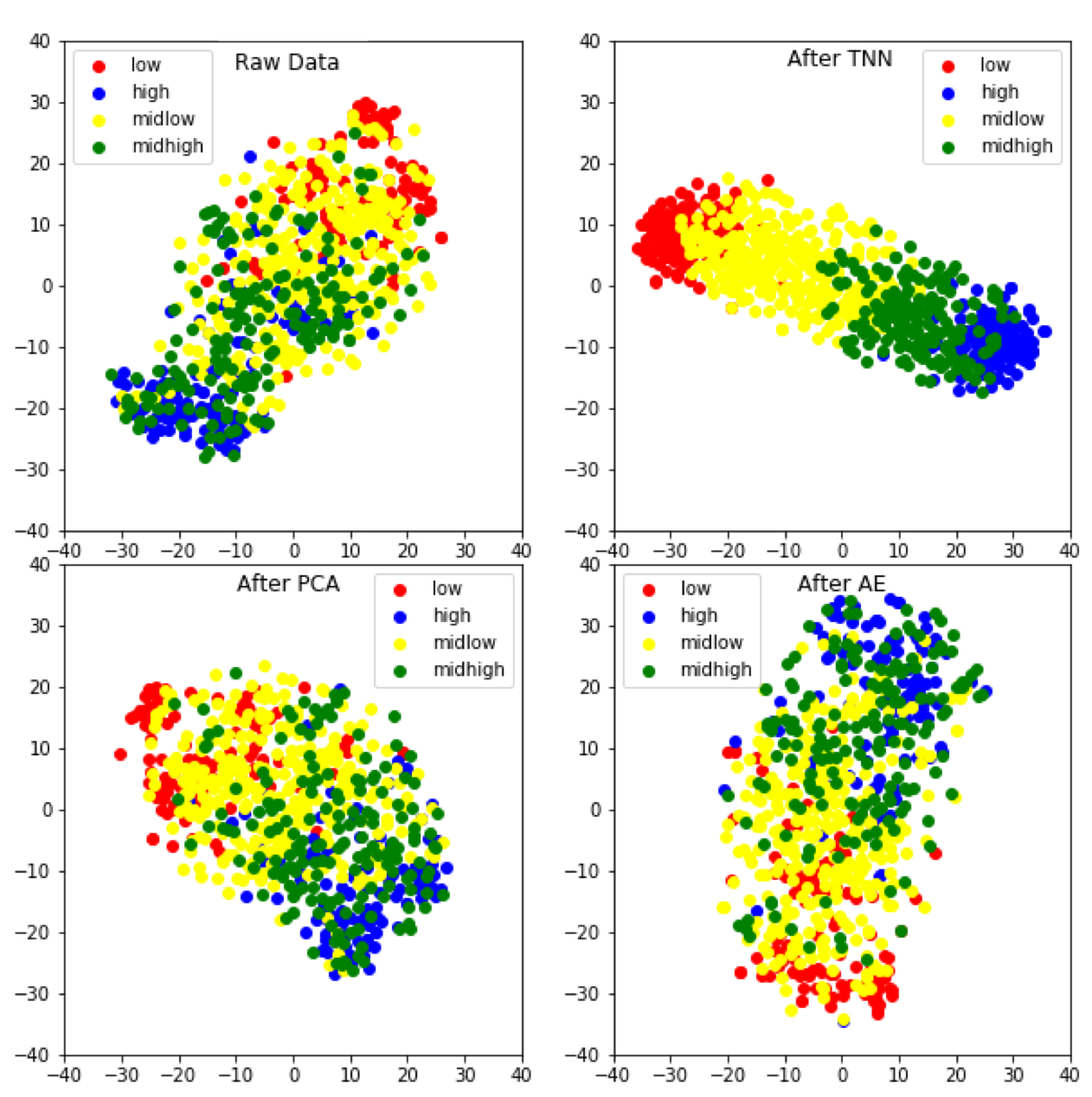}
	
	\caption{t-SNE visualization of the 6,669 MediaEval features and the 600 features after reduction by TNN, PCA, and AE.}
	% \label{fig:my_label}
	\label{fig:clusterplot}
	
\end{figure}

\section{Discussion}

\subsection{Visualization of learned embedded spaces.}
A visualisation of the data distribution before and after the TNN, PCA, and AE transformation is presented in Figure~\ref{fig:clusterplot}. A t-distributed stochastic neighbor embedding (t-SNE) was used to project the data into two-dimensions \cite{maaten2008visualizing}. We divided the songs into 4 classes for visualization, namely, high arousal, mid-high arousal, mid-low arousal, and low arousal values. The 100 songs with the highest arousal values were selected as the high arousal class, and the 100 songs with lowest arousal values as the low arousal class. We then split the remaining songs into mid-high and mid-low arousal classes using a similar procedure. The top left corner of Figure~\ref{fig:clusterplot} shows the data distribution when projecting 6,669 features into a two-dimensional plane by using t-SNE. The overlap between classes is considerable. The top right corner of Figure~\ref{fig:clusterplot} corresponds to the t-SNE projection on the TNN learned latent space for the data. It has more obvious clusters with only minor overlap. The bottom of the figure shows the data distribution when using PCA and AE for feature reduction. No clear clusters are formed under these two transformations.

\subsection{Result for the MediaEval experiment}
In our experiments, we study if we can still maintain a relatively good regression result, with reduced features learned by the TNN. From our experiment, we see that TNN-based models performed best when less layers were used. Therefore, we used a single layer fully connected network with ReLU activation as our TNN structure.
When comparing the results of our TNN method with other dimension reduction techniques (see Table \ref{tab:1}), We can see that the TNN has the ability to significantly reduce dimensionality of the data (more than 10\%, to 600 features), while still maintaining a relatively high $R^2$ score for both valence and arousal compared with the model using original features. Traditional methods such as PCA and RP were not able to deliver a similar performance with the reduced features. Although the performance of AE is marginally better than PCA, the resulting $R^2$ scores cannot match the performance of the TNN-SVR and TNN-GBM model. In order words, our proposed TNN method is more suited for reducing the dimension of the data in preparation of regression, when compared to traditional methods such as PCA, RP, and AE. We should note that TNN is a supervised clustering method as opposed to unclustered methods such as PCA, RP, and AE. Since we are assessing the potential of  different dimension reduction techniques as a precursor to perform a supervised classification problem, we argue that using the labels to perform the dimensionality reduction is a valid approach, as those labels are used for learning the final prediction.

The proposed TNN-SVR and TNN-GBM outperform \citet{markov2013music}'s model for both arousal and valence. Only in the case of valence, does the GPR~\cite{fukuyama2016music} reach a higher $R^2$ than our proposed model. It is important to note here that the number of features was reduced from 6,669 by one fold to 600, while still maintaining a relatively high $R^2$ score for both valence and arousal values. 
Readers should note that \citet{markov2013music} used seven fold cross-validation in their study instead of ten fold cross-validation. Since, the exact composition of the folds used for evaluation are unknown, it would therefore be worth exploring in future research if a TNN combined with exact set up as \cite{markov2013music} and \cite{fukuyama2016music} would achieve further improvements in performance. 

\begin{table}[htb]
	\small
	\centering
	\caption{Results for the MediaEval experiment: $R^2$ scores $\pm$ standard deviation for various models on the MediaEval 2013 dataset.}	
	\begin{tabular}{l|cc} %defining the number of columns
		\toprule
		& Valence & Arousal\\\midrule
		SVR\cite{markov2013music} & $0.112$ & $0.300$\\
		GPR\cite{markov2013music} & $0.170$ & $0.581$\\
		GPR \cite{fukuyama2016music} & $0.413\pm0.043$ & $0.636\pm0.040$\\\hdashline
		GBM(original features) &$0.431\pm0.089$ & $0.662\pm0.057$ \\
		PCA-GBM (600 features) & $0.251\pm0.118$ & $0.566\pm0.063$\\
		RP-GBM (600 features) & $0.229\pm0.152$ & $0.619\pm0.076$ \\
		AE-GBM (600 features) & $0.236\pm0.145$ & $0.578\pm0.077$\\
		TNN-GBM (600 features) &  \pmb{$0.374\pm0.058$} & \pmb{$0.621\pm0.080$} \\\hdashline
		SVR (original features) & $0.347\pm0.086$  & $0.614\pm0.054$ \\
		PCA-SVR (600 features) & $0.087\pm0.119$ & $0.224\pm0.140$\\
		RP-SVR (600 features) & $0.334\pm0.097$ & $0.608\pm0.062$\\
		AE-SVR (600 features) & $0.280\pm0.150$ & $0.598\pm0.100$ \\
		TNN-SVR (600 features)& \pmb{$0.378\pm0.066$} & \pmb{$0.638\pm0.055$}\\  
		\bottomrule
	\end{tabular}
 %with 6,669 features.} 
	%\dorien{Can you put the best values in bold? What is the plus/minus number? Explain in caption (same for other table)}}
	\label{tab:1}
	%\vspace{-4mm}%Put here to reduce too much white space after your table
\end{table}

\begin{table}[htb]
	\small
	\centering
	\caption{Results for the DEAM experiment: $R^2$ scores $\pm$ standard deviation for various models on the DEAM dataset.} %with 260 features.}
	\begin{tabular}{l|cc} %defining the number of columns
		\toprule
		& Valence & Arousal\\\midrule
		SVR (260 features) & $0.324\pm0.133$  & $0.638\pm0.062$ \\
		%TNN-SVR (260 features) & $0.366\pm0.124$ & $0.677\pm0.060$\\\hdashline
		GBM (260 features)& $0.318\pm0.151$ & $0.678\pm0.062$\\\hdashline
		
		PCA-GBM (100 features) & $0.288\pm0.123$ & $0.584\pm0.091$\\
		RP-GBM (100 features) & $0.251\pm0.137$ & $0.573\pm0.079$\\
		AE-GBM (100 features) & $0.312\pm0.109$ & $0.623\pm0.067$\\
		TNN-GBM (100 features) & \pmb{$0.367\pm0.113$} & \pmb{$0.662\pm0.065$} \\\hdashline
		PCA-GBM (50 features) & $0.251\pm0.124$ & $0.576\pm0.091$\\
		RP-GBM (50 features) & $0.215\pm0.149$ & $0.549\pm0.086$\\
		AE-GBM (50 features) & $0.255\pm0.146$ & $0.601\pm0.076$\\
		TNN-GBM (50 features)& \pmb{$0.339\pm0.124$} & \pmb{$0.661\pm0.068$}\\\hdashline
		
		PCA-SVR (100 features) & $0.274\pm0.121$ & $0.550\pm0.058$\\
		RP-SVR (100 features)　& $0.270\pm0.125$ & $0.615\pm0.061$\\
		AE-SVR (100 features) & $0.311\pm0.119$ & $0.629\pm0.070$\\
		TNN-SVR (100 features) & \pmb{$0.361\pm0.112$} & \pmb{$0.672\pm0.065$} \\\hdashline
		PCA-SVR (50 features) & $0.195\pm0.119$ & $0.444\pm0.063$\\
		RP-SVR (50 features) & $0.203\pm0.017$ & $0.574\pm0.077$ \\
		AE-SVR (50 features) & $0.269\pm0.159$ & $0.606\pm0.065$\\
		TNN-SVR (50 features)& \pmb{$0.352\pm0.112$} & \pmb{$0.669\pm0.070$}\\  
		\bottomrule
	\end{tabular}
	\label{tab:2}
\end{table}

\subsection{Result for the DEAM experiment} The experiments also show that GBM is more effective when dealing with the dataset that has a larger feature space (the MediaEval dataset), since it has a much higher $R^2$ score for both valence and arousal, compared to the GBM built on a reduced feature set. On the dataset with less features (the DEAM dataset), the TNN dimension reduction  marginally improves the GBM performance (see Table~\ref{tab:2}). SVR, on the other hand, is much more effective when dealing with features after dimension reduction. When using SVR, the TNN can achieve a 90\% feature reduction with a 9\% improvement in valence prediction $R^2$ and 4\% improvement in arousal prediction $R^2$. When using GBM, the TNN can achieve the same feature reduction with 13\% decrease in valence prediction $R^2$ and 6\% decrease in arousal prediction $R^2$. For both regression algorithms, the TNN outperforms all other traditional dimensionality reduction algorithms such as PCA, RP, and AE.

We show that TNNs can further reduce the 260 DEAM features to only 100 or 50 features without a huge impact on the regression accuracy. In the case of SVR with reduced TNN features, the $R^2$ scores for both valence and arousal still outperform the baseline SVR model. Among PCA, RP, and AE, PCA performs the worst, probably because it is a linear process, while the other two can capture non-linearity. When the number of features is reduced from 260 to 50, PCA-SVR has a $40\%$ decrease in $R^2$ scores for valence, and a $30\%$ decrease for arousal. Although AE with either SVR or GBM performs better than PCA, it still has a $17\%$ and $5\%$ decrease in valence and arousal $R^2$ scores when the dimension is reduced to 50. TNNs, on the other hand, are able to better capture the non-linear relations among the original features, thus, successfully reducing the features while improving the valence and arousal prediction $R^2$ score by $8\%$ and $5\%$ respectively. A similar trend can be observed for the GBM-TNN case, despite a slight $2.5\%$ decrease in $R^2$ score for arousal value when the TNN's layer size is reduced to 50.

\section{Conclusion}
\label{sec:conclusion}

%We propose an efficient strategy for dimension reduction based on triplet neural networks for an audio-based regression problem, namely, music emotion prediction. Since TNNs are most often used in a classification context, 
We  propose a strategy to leverage triplet neural networks for regression tasks with a new adaptative mining of negative and positive samples, And we show its efficiency on music emotion regression. Based on two experiments (on the MediaEval and DEAM datasets), we see that our hybrid TNN, combined with a SVR or GBM regressor, has the ability to perform significant dimensionality reduction while still improving the $R^2$ scores. Traditional methods such as principal component analysis, and autoencoders failed to maintain the same level of accuracy. We believe that this paper provides a foundation for a deeper and more comprehensive future study of applying TNNs to regression problems. Some of the open questions that we aim to further investigate are: (1) Can our TNN reduce the latent space to only 3 or 2 components? Why do spectrogram features result in a less effective TNN latent space than the openSMILE features? 

\section*{Acknowledgment}
We would like to thank the anonymous reviewers for their constructive reviews. This work is supported by a Singapore International Graduate Award (SINGA) provided by the Agency for Science, Technology and Research (A*STAR), under grant no. SING-2018-02-0204. Moreover, the equipment required to run the experiments is supported by SUTD-MIT IDC grant no. IDG31800103 and MOE Grant no. MOE2018-T2-2-161.

\bibliography{refs}
\bibliographystyle{IEEEtranN}

\end{document}